\documentclass[oldversion]{aa}  
\usepackage{natbib}
\usepackage{graphicx}
\usepackage{rotating}
\usepackage{txfonts}
\newcommand{\ms}{$\,$M$_\mathrm{\odot}$}

\newcommand{\Teff}{$\,$T$_\mathrm{eff}$}
\newcommand{\logg}{$\,\log$~g}
\newcommand{\be}{\begin{equation}}
\newcommand{\ee}{\end{equation}}
\newcommand{\stars}{{\sc stars}}
\newcommand{\el}[2]{\ensuremath{^{#1}\mathrm{#2}}}

\newcommand{\pg}{\ensuremath{\mathrm{(p,\gamma)}}}

\newcommand{\gr}{\bigtriangledown_\mathrm{rad}}
\newcommand{\gad}{\bigtriangledown_\mathrm{ad}}

\newcommand{\deltaov}{\ensuremath{\delta_\mathrm{ov}}}

\begin{document}
   \title{Confronting uncertainties in stellar physics \\ II. exploring differences in main-sequence stellar evolution tracks}
	\titlerunning{CUSP II -- stellar tracks}

   \author{R. J. Stancliffe\inst{1} \and L. Fossati\inst{2,1} \and J.-C. Passy\inst{1} \and F. R. N. Schneider\inst{1,3}}

   \institute{Argelander-Institut f\"ur Astronomie, University of Bonn, Auf dem H\"ugel 71, D-53121 Bonn, Germany
   \and Space Research Institute, Austrian Academy of Sciences, Schmiedlstrasse 6, A-8042 Graz, Austria
   \and Department of Physics, University of Oxford, Denys Wilkinson Building, Keble Road, Oxford OX1 3RH, U.K.}

   \date{}

\abstract{We assess the systematic uncertainties in stellar evolutionary calculations for low- to intermediate-mass, main-sequence stars. We compare published stellar tracks from several different evolution codes with our own tracks computed using the stellar codes \stars\ and {\sc mesa}. In particular, we focus on tracks of 1 and 3\ms\ at solar metallicity. We find that the spread in the available 1\ms\ tracks (computed before the recent solar composition revision by Asplund et al.) can be covered by tracks between 0.97-1.01\ms\ computed with the \stars\ code. We assess some possible causes of the origin of this uncertainty, including how the choice of input physics and the solar constraints used to perform the solar calibration affect the tracks. We find that for a 1\ms\ track, uncertainties of around 10\% in the initial hydrogen abundance and initial metallicity produce around a 2\% error in mass. For the 3\ms\ tracks, there is very little difference between the tracks from the various different stellar codes. The main difference comes in the extent of the main sequence, which we believe results from the different choices of the implementation of convective overshooting in the core. Uncertainties in the initial abundances lead to a 1-2\% error in the mass determination. These uncertainties cover only part of the total error budget, which should also include uncertainties in the input physics (e.g., reaction rates, opacities, convective models) and any missing physics (e.g., radiative levitation, rotation, magnetic fields). Uncertainties in stellar surface properties such as luminosity and effective temperature will further reduce the accuracy of any potential mass determinations.}

   \maketitle
%

\section{Introduction}

Stellar evolution codes are a lot like religions: there are many of them to choose from, they possess many similarities, and it is not obvious which of them (if any) is correct. The outputs of these codes, in the form of evolutionary tracks and isochrones, are used in several fields of astrophysics, ranging from the study of extra-solar planets (exoplanets) to stellar populations. Many grids of tracks and isochrones, computed with different stellar evolution codes, are freely available and are used to derive fundamental properties of stars (e.g., mass, radius, age) and star clusters (e.g., age, distance) on the basis of observables (e.g., magnitudes, effective temperature, surface gravity, luminosity, chemical composition). When comparing observations to stellar evolution tracks one often considers only the observational uncertainties in the error budget. This neglects the systematic theoretical uncertainties in the stellar codes (see below), some of which are difficult to quantify.

In the exoplanet field, evolutionary tracks are commonly used to derive the mass and radius of planet-hosting stars on the basis of properties measured from spectra and transit light curves (e.g., effective temperature, surface gravity, and average density). The stellar parameters are then used to directly derive the planet mass and radius, and age of the system; as a consequence, systematic uncertainties in the stellar parameters propagate to the planet parameters as well. It is important to note that small systematic uncertainties in the stellar parameters may be strongly significant in terms of planet properties, for example for planets in the 5--10\,M$_\oplus$ transition region between gaseous and rocky planets. For these planets underestimating the uncertainty of their mass and/or radius may lead to dramatically wrong conclusions: a planet presumed to be a super-Earth in the habitable zone may instead by an uninhabitable gaseous planet. This is particularly relevant because many planet-finding facilities, which are about to begin their operations (e.g., NGTS [\citealt{2013EPJWC..4713002W}], TESS [\citealt{2014SPIE.9143E..20R}], and CHEOPS [\citealt{2013EPJWC..4703005B}]), are designed to have their maximum efficiency in finding mini-Neptunes and super-Earths orbiting nearby late-type stars.

With the ``Confronting Uncertainties in Stellar Physics'' (CUSP) project we intend to estimate the systematic uncertainties involved in the calculation of stellar evolution tracks, for example to evaluate their impact in terms of planet parameters. For this task, we will use {\sc bonnsai}\footnote{The {\sc bonnsai} web-service is available at {\tt www.astro.uni-bonn.de/stars/bonnsai}.} \citep{2014arXiv1408.3409S}, which, to the best of our knowledge, is the only publicly available tool allowing one to derive stellar parameters (e.g., mass, radius, age) from a set of observational parameters (e.g., effective temperature, surface gravity, luminosity), using Bayesian statistics to properly account for the observational uncertainties. In its current status, {\sc bonnsai} contains evolutionary tracks only for massive stars (M$>$5\,M$_\odot$; \citealt{2011A&A...530A.115B, 2015A&A...573A..71K}). We wish to extend the model database to lower mass stars from 0.8 to 10\,M$_\odot$ on the basis of grids calculated using the \stars\ \citep{1971MNRAS.151..351E,2009MNRAS.396.1699S} and {\sc mesa} \citep{2011ApJS..192....3P,2013ApJS..208....4P,2015arXiv150603146P} stellar evolution codes.

In the first paper of this series \citep{2015A&A...575A.117S}, we used a set of 12 eclipsing binaries, with component masses between 1.3 and 6.2\,M$_\odot$, to constrain the convective core overshooting parameter to be adopted for the model grids. We concluded that the majority of the analysed systems can be well described with a mass-independent overshooting parameter corresponding to an extension of the mixed region above the core of about 0.1--0.3 pressure scale heights. In this work, we aim at increasing the awareness of users (observers in particular) of grids of stellar evolution models and isochrones on the presence of theoretical uncertainties and systematics. We also highlight caveats present when converting stellar atmospheric parameters (i.e., effective temperature, surface gravity, and metallicity) into fundamental parameters (i.e., mass, radius, and age). We do this here by investigating the uncertainties in main-sequence, low-mass stellar evolution tracks, based on comparisons between tracks for 1 and 3\ms\ models (comparisons for high-mass stars have recently be carried out by \citealt{2013A&A...560A..16M} and \citealt{2015MNRAS.447.3115J}). We limit ourself to the main sequence phase of evolution, because it is the one which is most intensively considered in terms of stellar spectroscopic analysis and conversion of the atmospheric parameters into stellar mass, radius, and age (e.g., the vast majority of the known planet-hosting stars are on the main sequence). We will consider uncertainties in later stages of evolution in future work.

Uncertainties in stellar modelling can be divided into two groups: physical uncertainties and computational/numerical uncertainties. The former group includes uncertainties in measured quantities used as inputs for computations, such as reaction rates and opacities. It also includes our ignorance of physical processes such as the efficiency of mixing processes and the nature of mixing at convective boundaries \citep{2015arXiv150603100V}. The second group relates to how the equations of stellar structure and evolution are solved. It includes issues related to spatial and temporal resolution in models, the use of simultaneous or non-simultaneous solution of the equations of stellar structure \citep{2006MNRAS.370.1817S}, the choice of solution scheme \citep{2004A&A...416.1023M}, and the algorithmic  treatment of mixing at convective-radiative boundaries \citep{2014A&A...569A..63G} to name a few examples.

We have collected stellar tracks from various publicly available sources, as well as computing tracks with two other codes. This is by no means an exhaustive list of what is available, but we believe it provides a representative range of current stellar evolution calculations. The codes used are:
\begin{itemize}
\item{{\sc mesa}. This code is described extensively by \citet{2011ApJS..192....3P}, \citet{2013ApJS..208....4P} and \cite{2015arXiv150603146P}. We use revision 7503 of the code.}
\item{\stars. The \stars\ stellar evolution code was originally developed by \citet{1971MNRAS.151..351E} and has been updated by many authors \citep[e.g.][]{1995MNRAS.274..964P}. The code is freely available for download{\footnote{\texttt{http://www.ast.cam.ac.uk/$\sim$stars}}}. This code solves the equations of stellar structure and chemical evolution in a fully simultaneous manner, iterating on all variables at the same time in order to converge a model \citep[see][for a detailed discussion]{2006MNRAS.370.1817S}. The version employed here is that of \citet{2009MNRAS.396.1699S} which was developed for doing binary stellar evolution. The code treats all forms of mixing by means of a diffusive formalism \citep{1972MNRAS.156..361E}.}
\item{Y$^2$. Tracks from the Yonsei-Yale collaboration are available online\footnote{\texttt{http://www.astro.yale.edu/demarque/yystar.html}}. Details of the code and its calibration can be found in \citet{2001ApJS..136..417Y}.}
\item{BASTI. Tracks from BASTI are taken from the BASTI website\footnote{\texttt{http://basti.oa-teramo.inaf.it/index.html}}. Details of the code are described in \citet{2004ApJ...612..168P}.}
\item{Padova. The Padova tracks have been taken from their website\footnote{\texttt{http://stev.oapd.inaf.it/YZVAR/}}. Details of the code and the models are described in \citet{2008A&A...484..815B}.}
\item{Dartmouth. The Darmouth stellar evolution code is described in \citet{2006ApJ...641.1102B} and \citet{2007AJ....134..376D}. The models used here come from the work of \citet{2008ApJS..178...89D} and are available online\footnote{\texttt{http://stellar.dartmouth.edu/models/index.html}}.}
\item{Geneva. These models are taken from the work of \citet{2012A&A...541A..41M} and are available from the Geneva website\footnote{\texttt{http://obswww.unige.ch/Recherche/evol/-Database-}}.}
\item{PARSEC. This is an updated version of the Padova code and is described in \citet{2012MNRAS.427..127B}. Tracks may be obtained online\footnote{\texttt{http://stev.oapd.inaf.it/cgi-bin/cmd}}.}
\item{Victoria-Regina. This code is described in \citet{2012ApJ...755...15V} and references therein. The tracks present here come from the work of \citet{2014ApJ...794...72V}, and are available for download\footnote{\texttt{http://www.canfar.phys.uvic.ca/vosui/\#/VRmodels}}.}
\end{itemize}
Many of these codes and grids are massively used, mostly by observers, to convert spectroscopically derived stellar atmospheric parameters into fundamental parameters. This conversion is often done without considering the physics included by the different grids, whether the adopted input parameters (e.g., reaction rates) are up to date, the calibration procedures, and the true meaning of some of the basic parameters (e.g., present-day vs initial metallicity). Given the large amount and quality of observational data already available, for example from large spectroscopic surveys, and soon to come, for example from GAIA, it is imperative that the different communities are fully aware of the presence and size of the systematic uncertainties associated with stellar evolution calculations.

It should be stressed that this paper is not intended to represent a code comparison, that is to say, a test of the {\it mechanics} of the various stellar evolution codes. Such comparisons have been performed by many groups (e.g. at The Aarhus Red Giants Workshop\footnote{\texttt{http://users-phys.au.dk/victor/rgwork/}}). The most important aspect of a code comparison is to completely standardize the input physics and then compare the outputs, and thus assess whether the machinery of the stellar codes works the same way. However, for the various published tracks produced by different groups the choices of input physics is {\it not} the same. At some point, choices have to be made (for example, for the inclusion of diffusion as will be discussed below) and it is these choices that lead to the differences between the calculations made by different groups. Although we are making no assertions in this work as to which choices are correct, users of published stellar models should critically examine how well the  available grids satisfy empirical constraints (e.g. the solar calibration, binary stars) and the extent to which they take into account recent advances in stellar physics (e.g. updated reaction rates, opacities). This represents (some of) the theoretical uncertainties inherent in stellar modelling. Our aim is to put ourselves in the perspective of someone who needs to derive e.g., the mass of a star on the basis of spectroscopically-derived atmospheric parameters, and to estimate the uncertainties involved in this process caused simply by choosing one set of tracks instead of another. We will address the issue of uncertainties in ages in a separate paper.

\section{Comparison of 1\ms\ tracks}

\begin{figure}
\includegraphics[width=\columnwidth]{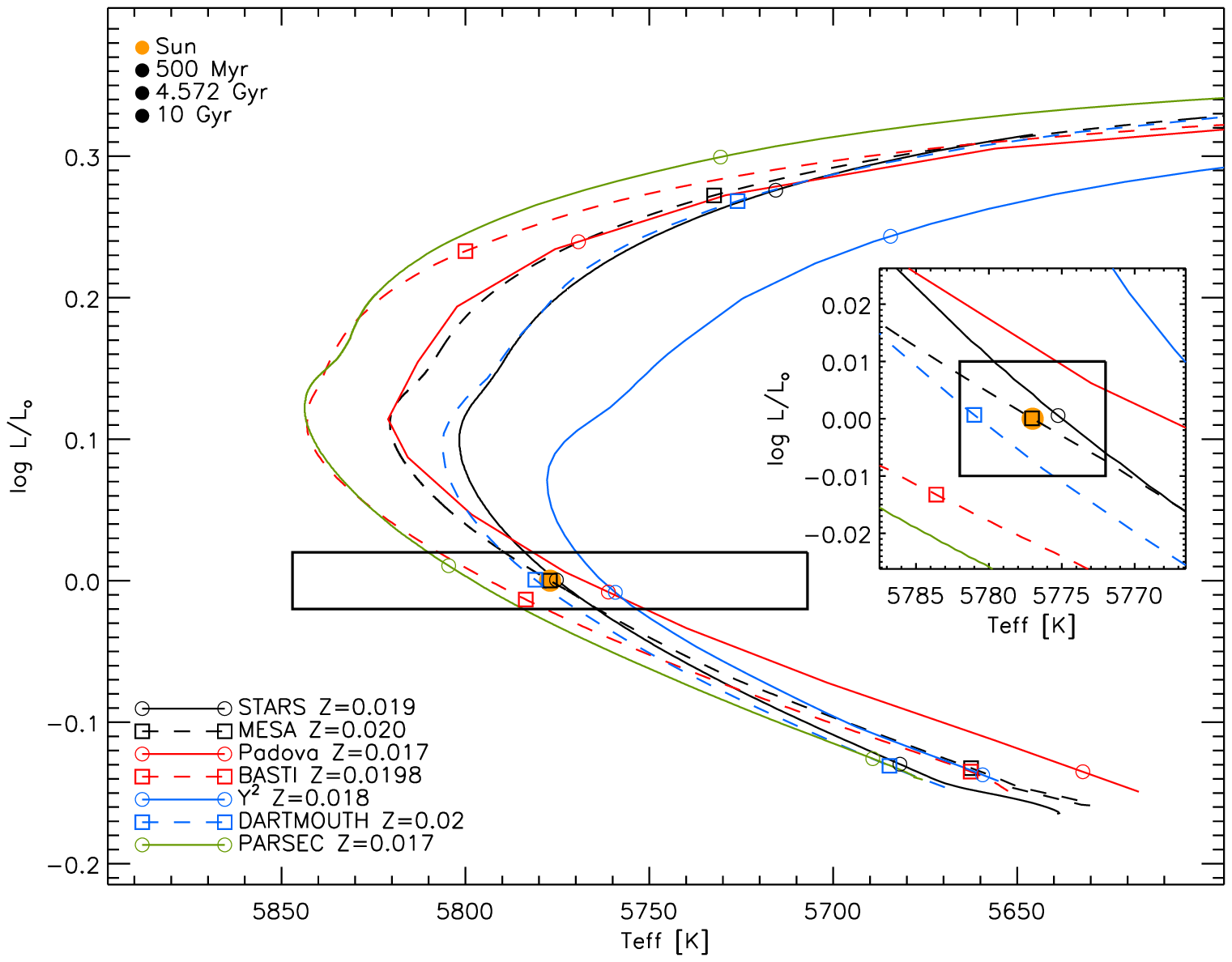}
\includegraphics[width=\columnwidth]{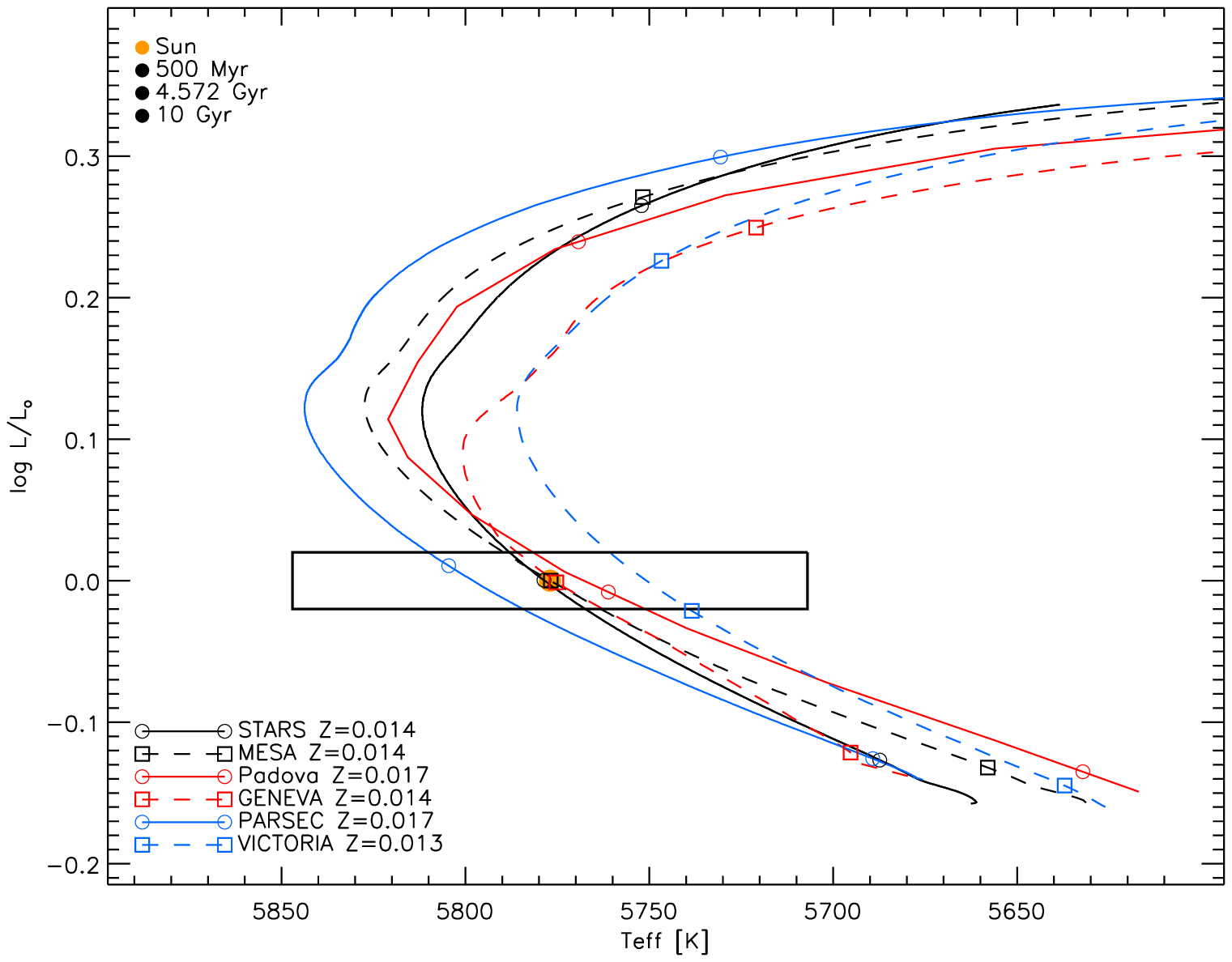}
\caption{Evolutionary tracks for a 1\ms\ model for various stellar evolution codes. The box represents the estimated post-GAIA uncertainties of the luminosity and effective temperature. {\bf Top:} Models comuputed with a metallicity of Z\,$\approx0.02$. The inset shows a zoomed in region around the solar luminosity and temperature, with the box displaying the uncertainties of the solar luminosity and effective temperature. {\bf Bottom:} Models computing using the abundances of \citet{2009ARA&A..47..481A}. The Padova track, which is intermediate in metallicity is included in both panels.}
\label{fig:1sm}
\end{figure}

In Fig.~\ref{fig:1sm}, we plot evolutionary tracks for a 1\ms\ model computed with the evolution codes listed above. We have separated the tracks into two groups based on their metallicity: those plotted with an `old' solar metallicity are shown in the upper panel, while those with the \citet{2009ARA&A..47..481A} metallicity are plotted in the lower panel. As a shorthand, we will use the term `old' to refer to pre-\citet{2009ARA&A..47..481A} solar models, and `new' to refer to models computed with the \citet{2009ARA&A..47..481A} abundances. Perfect agreement with the Sun should not be expected as we are plotting 1\ms\ evolutionary tracks computed for grids of stellar models which are not necessarily the solar calibration tracks themselves. The exceptions here are {\sc stars} and {\sc mesa}, as these are our own solar-calibration tracks. In some cases, e.g. BASTI, the solar calibration is performed with diffusion, but the publicly available track is computed without. For the `old' Sun, most of the codes give effective temperatures within 20-30\,K of the solar value of 5777 K \citep{1996Sci...272.1286C}. In all cases, luminosity variations are extremely small. The box drawn in Fig. 1 shows what we consider to be typical post-GAIA uncertainties. The error bar on the luminosity has been derived assuming that the whole uncertainty is in the bolometric correction (0.05\,mag; see e.g., \citealt{1996ApJ...469..355F}). The uncertainty on the effective temperature is the average one typically expected from state-of-the-art analysis of high quality, high resolution spectra of solar-like stars \citep[70\,K; see e.g.][]{2015arXiv151106134R}. All these tracks would lie comfortably within the estimated post-GAIA errors. The spread in temperatures between the tracks at the solar age is comparable to the initial spread in temperatures at 500\,Myr, which is around 80\,K. At the 10 Gyr point, the spread in temperatures is much larger, around 150\,K which is about the size of the whole error box. The spread in luminosity remains small at around 0.03 dex. The spread in the stellar tracks represents the uncertainty in the calculation of a 1\ms\ star. The hottest track is roughly reproduced by a 1.01\ms\ \stars\ model, while the coolest requires a 0.97\ms\ model. Thus if we were to observe a star with the Sun's luminosity and temperature (and metallicity), the error on the mass we determine by fitting a 1\ms\ track to it would be around $\pm0.02$\ms.

Much of the spread between the various `old' solar tracks comes from the variety of input physics. We list some of these in Table~\ref{tab:solar}. Not everyone calibrates the Sun using the same physics. There are a variety of choices of present-day solar abundances, which affects the bulk metallicity and initial hydrogen mass fraction. Gravitational settling/diffusion\footnote{We shall use these two terms interchangeably.} of elements is considered by many, but not all codes. On top of this, not everyone uses the same set of constraints for the calibration. In addition to using the current solar luminosity and radius, some codes also use the depth of the solar convection zone. We shall discuss the effects of these different choices below. In the case of the `new' solar model, the agreement between codes is far better (see the lower panel of Fig.~\ref{fig:1sm}) presumably as a result of a much more homogenous set of input physics: all the codes use very similar abundances and all include the effects of diffusion, although the Victoria-Regina code only applies diffusion to helium.

\begin{table}
\begin{center}
\begin{tabular}{lccccc}
\hline
Code & solar & Z$_0$ & X$_\mathrm{0}$ & $\alpha_\mathrm{MLT}$ & diffusion \\
& abundance & \\
\hline
\stars & GN93 & 0.0184 & 0.7012 & 2.090  & yes \\
	& A09 & 0.0142 & 0.7293 & 2.025 & yes \\
{\sc mesa} & GS98 & 0.0185 & 0.7101  & 1.877 & yes \\
	& A09 & 0.0149 & 0.7193 & 1.783 & yes \\
Y$^2$ & GNS96 & 0.0181 & 0.7149 & 1.7432 & He only \\
BASTI & GN93 & 0.198  & 0.7068 & 1.913 & yes \\
Padova & GS98 & 0.017 & 0.723 & 1.68 & no \\
Dartmouth & GS98 & 0.0188 & 0.7063 & 1.938 & yes\\
Geneva & A05 & 0.014 & 0.7200  & 1.6467 & yes \\
PARSEC & GS98+C11 & 0.0177 & 0.7027  & 1.74 & yes \\
Victoria- & A09 & 0.0133 & 0.7367 & 2.007 & He only \\
Regina & \\
\hline 
\end{tabular}
\end{center}
\caption{Details of the parameters for the solar calibrations of the various codes included in this study. The abbreviations for the choice of solar abundances are: GN93, \citet{1993oee..conf...15G}; GNS96, \citet{1996ASPC...99..117G}; GS98, \citet{1998SSRv...85..161G}; A05, \citet{2005ASPC..336...25A}; A09, \citet{2009ARA&A..47..481A}; C11, \citet{2011SoPh..268..255C}. Note that the quoted values for {\sc mesa} are based on the calibration with $L, T_\mathrm{eff}$ and $Z/X$ alone.}
\label{tab:solar}
\end{table}

The inclusion of diffusion in stellar codes presents some issues. From a purely physical perspective, diffusion is a process that should be included in stellar calculations. \citet{1993ApJ...403L..75C} showed that the inclusion of helium settling has a significant effect on the structure and oscillation frequencies of the Sun. However, diffusion alone is too efficient at depleting materials and an additional mixing process seems to be necessary. \citet{1999ApJ...525.1032B} showed that for the Sun, turbulent mixing can inhibit settling by 25\% and that this leads to a significant improvement in the solar sound speed profile. At low metallicity, diffusion together with turbulent mixing may present an explanation for the uniform Li abundances observed in unevolved stars (the so-called Spite plateau) which solves the cosmic lithium problem without the need for any exotic physics during Big Bang nucleosynthesis \citep[see][and references therein]{2005ApJ...619..538R}. A similar conclusion is reached by \citet{2006Natur.442..657K} from the observation of lithium in the metal-poor globular cluster NGC~6397. The behaviour of heavier elements in this same cluster is also consistent with stellar models including diffusion and turbulent mixing \citep{2008A&A...490..777L}. In addition to these surface effects, the inclusion of diffusion may reduce the age of a low-mass star at a given turn-off luminosity \citep{1991ApJS...77..473P} and is therefore an important consideration if one wishes to determine stellar ages.

It therefore seems that one should prefer stellar tracks computed with the inclusion of diffusion. However, there is an important caveat. First, the above evidence points to the need for more than just diffusion alone. An additional mixing processes is required to inhibit the effects of diffusion. No physical cause for this mixing has been identified and so the prescriptions used are entirely {\it ad hoc}. Additionally, most of the codes presented here do not employ such turbulent mixing. Both the Dartmouth and PARSEC codes inhibit the action of diffusion in an ad hoc way, but with different prescriptions. The Victoria-Regina code adds a turbulent diffusion coefficient which is similar in form to the \citet{2005ApJ...619..538R} prescription \citep{2012ApJ...755...15V}. None of the codes describe here includes the effects of radiative levitation, which is the counterpart of gravitational settling, though it typically only influences the heavier elements \citep{2002ApJ...580.1100R}. There is clearly more to the issue of settling than we currently understand. For this reason, we do not dismiss codes that do not include it. We shall see below the potential error that may be introduced by not including diffusion.

Diffusion also provides the observer with another headache: which composition should the stellar tracks used for comparison have? Theorists label their tracks by initial abundance, but an observer sees only the present-day surface abundance. If settling has substantially altered the latter, it no longer reflects the former. For the Sun, the current surface metallicity is thought to be around 10\% lower than the initial metallicity \citep{2009ARA&A..47..481A}. We are able to estimate this because we know the solar age and hence we know the Sun's evolutionary state. For a given field star, this information is not available and we have to live with the consequences of our ignorance.

\subsection{The effects of the choice of solar calibration}

To calibrate stellar evolutionary tracks, it is typical to calibrate a code's mixing length parameter using a single reference point, namely the Sun. It is well known that with the current available input physics and the \citet{2009ARA&A..47..481A} abundances a 1\ms\ star at the current age of the Solar System does not give a structure consistent with helioseismic measurements \citep[see e.g.][and references therein]{2014ApJ...787...13V}. Specifically, one cannot produce a model that fits the current radius and luminosity of the Sun which also has the correct depth of the convective envelope.

\begin{figure}
\includegraphics[width=\columnwidth]{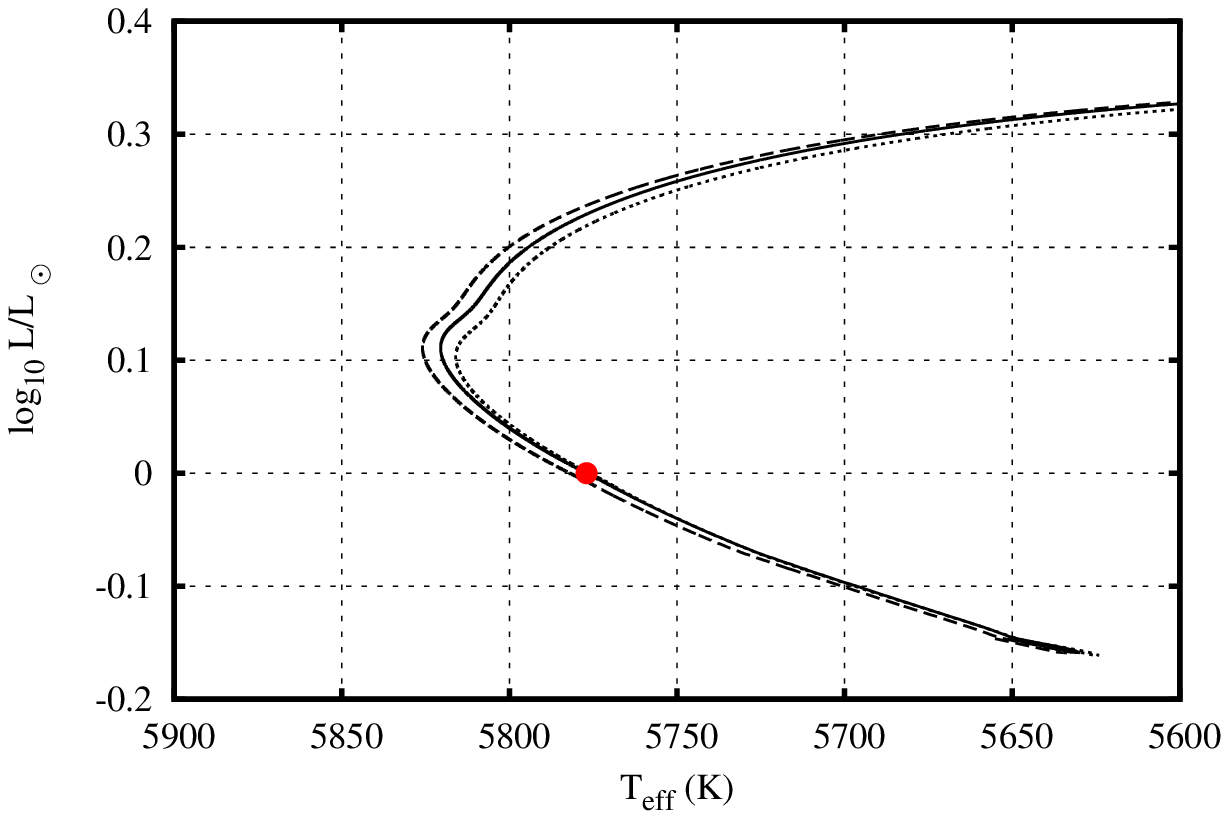}
\includegraphics[width=\columnwidth]{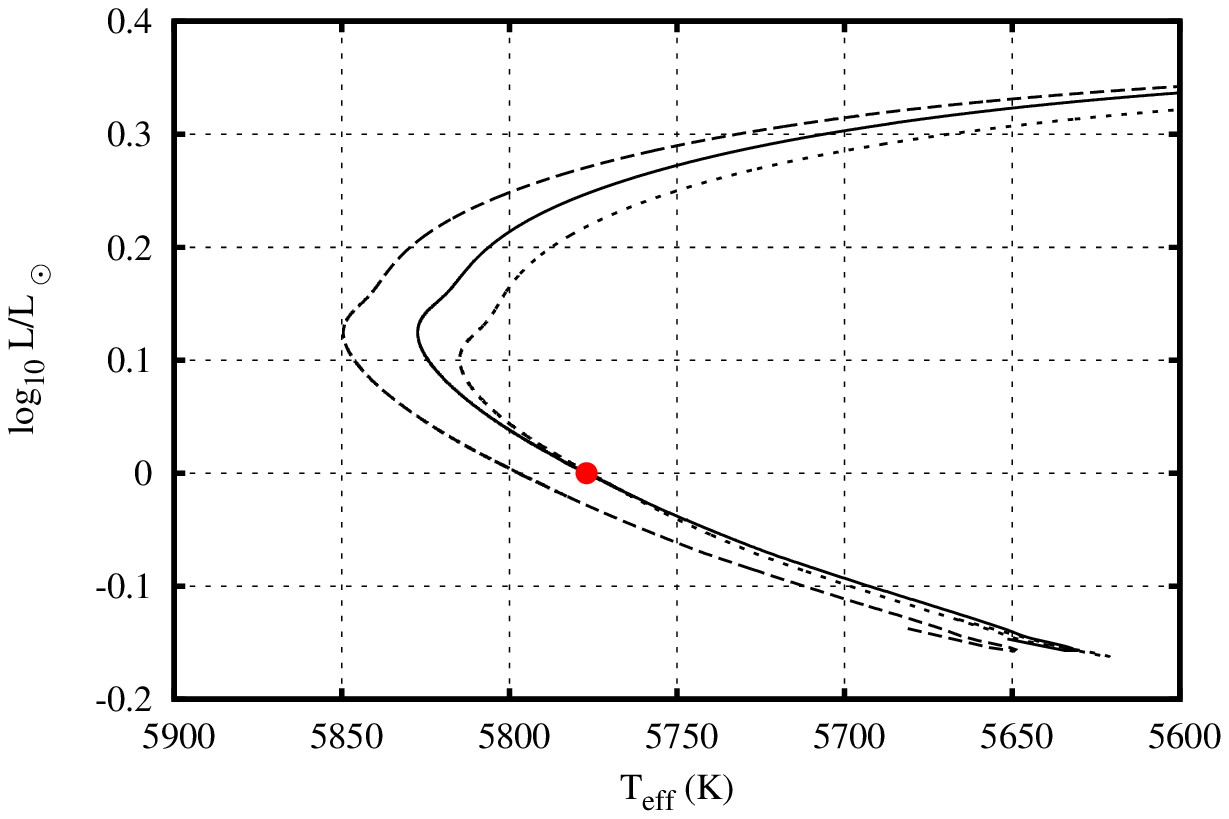}
\caption{The effect of different calibrations for the Sun using {\sc MESA}. The solid line indicates the track for  calibration performed using just the solar luminosity, temperature and Z/X. The dashed line additionally includes the depth of the solar convection zone. The dotted line includes all constraints including the RMS sound speed. The red point represents the Sun. The upper panel shows tracks computed with the GS98 abundances, while the lower one shows tracks computed with the A09 abundances.}
\label{fig:MESA_calibration}
\end{figure}

To test the effects of the choice of calibration constraints, we perform the following tests using the {\sc mesa} code. We first calibrate the solar model using only the solar luminosity, temperature and Z/X. We then repeat the calibration adding additional constraints. First, we add the depth of the solar convection zone $R_{\mathrm{cz}}$, followed by the helium abundance in the solar convection zone. Lastly, we consider the RMS deviation of the model's sound speed profile $\delta c_s$ from that deduced from helioseismology \citep{1997SoPh..175..287R}. The values used are given in Table~\ref{tab:observables}. In each case a $\chi^2$ minimisation is used to determine the best fit model, i.e.
\begin{equation}
	\chi^2 = \frac{1}{n}\sum_{i} \left( \frac{\bar{X_i} - \tilde{X_i}}{\sigma_{X_i}} \right)^2
\end{equation}
where $n$ is the number of observables, $X_i$ is a given observable, $\bar{X_i}$ its value for the model, $\tilde{X_i}$ its observed value, and $\sigma_{X_i}$ the observed uncertainty. This is evaluated at the solar age of 4.57\,Gyrs \citep{1995RvMP...67..781B}. We use the simplex algorithm \citep{NelderMead} implemented in {\sc mesa} to minimize the $\chi^2$ and stop the calibration once changes in the initial parameters no longer affect the obtained $\chi^2$. The code is free to change the initial metallicity, hydrogen abundance and mixing length parameter. We perform this calibration for both the \citet{1998SSRv...85..161G} and the \citet{2009ARA&A..47..481A} solar abundances, using OP opacities \citep{2005MNRAS.360..458B} and including gravitational settling. The results of this calibration are shown in Fig.~\ref{fig:MESA_calibration} and the details of the best fit models are presented in table~\ref{tab:mesa_solar}.

\begin{table}
\begin{center}
\begin{tabular}{cccl}
\hline
$X$ & $\tilde{X}$ & $\sigma_X$ & Reference \\
\hline
$\log{L/L_\odot}$ & 0.0 & $4\times10^{-4}$ & \citet{2004AARv..12..273F}\\
$T_{\mathrm{eff}}$ & 5777 K & 5 K & \citet{2015AA...573A..25S} \\
Z/X & 0.0231 or 0.0181 & 0.002 & GS98, A09\\
He & 0.2485 & 0.0034 & \citet{2004ApJ...606L..85B} \\
$R_{\mathrm{cz}} / R_\odot $ & 0.713  & $10^{-3}$ & \citet{1997MNRAS.287..189B} \\
$\delta c_s$ & 0.0 & $10^{-4}$ & \citet{1997SoPh..175..287R} \\
\hline
\end{tabular}
\end{center}
\caption{Observed values used for the {\sc mesa} calibration. The value of Z/X depends on the use of the GS98 (0.0231) or A09 (0.0181) abundances.}
\label{tab:observables}
\end{table}

For the GS98 abundances, the calibrations all give extremely similar results. This is not particularly surprising -- it is well established that the `old' solar abundances were extremely well reconciled with helioseismic measurements. The exact choice of calibration should not be crucial when using `old' solar abundances. For the `new' Sun, the choice of calibration constraints has a noticeable effect. When the depth of the solar convection zone is added, the track no longer has a luminosity and effective temperature that fit the Sun: the model is 20\,K too hot and has $\log L/L_\odot = -4\times10^{-4}$ (i.e. about 0.1\% below the solar luminosity). When we include the sound speed profile in the calibration, the fit in the HR diagram is much improved. However, this is achieved at the expense of the surface Z/X, which is 0.0246 and thus clearly incompatible with the \citet{2009ARA&A..47..481A} abundances. This model cannot be considered a reasonable calibration.

\subsubsection{Settling}

\begin{figure}
\includegraphics[width=\columnwidth]{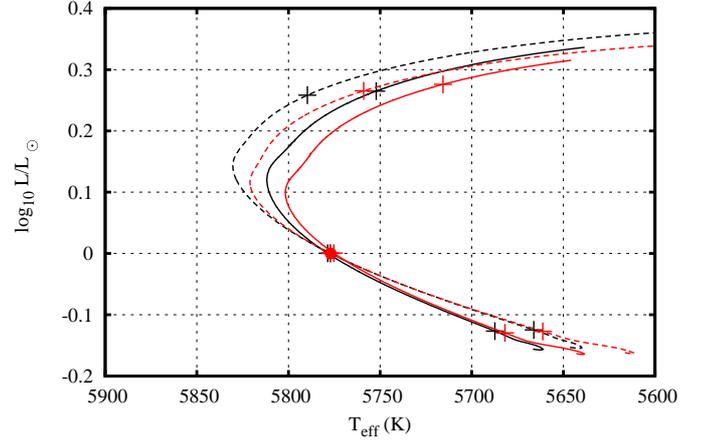}
\caption{Calibrated evolutionary tracks for the `old' (red) and `new' (black) Sun. The solid lines were computed using gravitational settling while the dashed lines were computed without. Pluses mark ages of 500\,Myr, 4.57\,Gyr and 10\,Gyr. The red circle represents the location of the Sun.}
\label{fig:Settling}
\end{figure}

Here we illustrate the difference between models calibrated with gravitational settling and those without. In Fig.~\ref{fig:Settling}, we show 1\ms\ tracks for both the `old' and `new' Sun computed with and without settling. These tracks have been calculated using the \stars\ code. The calibration for the models without diffusion (Z$_0$, X$_0$, $\alpha_\mathrm{MLT}$) give (0.0132, 0.7279, 1.900) for the `new' Sun and (0.0171, 0.7000, 1.965) for the `old' Sun. The calibrations have more of an effect on temperature than they do on luminosity. Close to the ZAMS, the effective temperatures of the four models differ by about 20\,K. By 10\,Gyr, this spread has grown to around 80\,K. At the hottest point, the spread in temperature is around 30\,K. Prior to the solar calibration point, models without gravitational settling are cooler than those that include settling. Once the calibration point is passed, the situation is reversed and the models without settling are hotter than those with settling.

\begin{figure}
\includegraphics[width=\columnwidth]{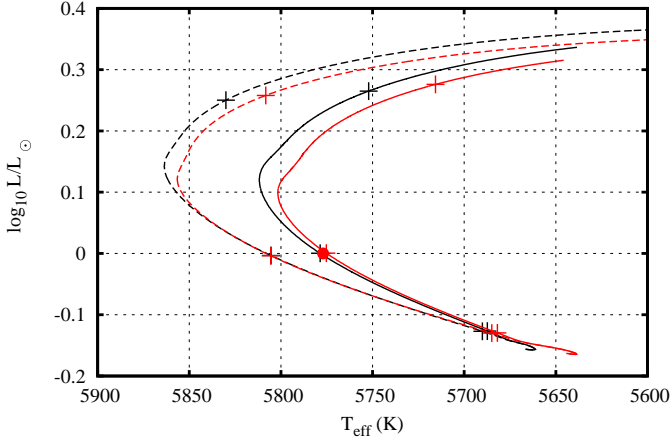}
\caption{Evolutionary tracks for the `old' (red) and `new' (black) Sun for the initial conditions listed in Table~\ref{tab:solar}. The solid lines were computed using gravitational settling while the dashed lines were computed without. The red circle represents the location of the Sun.}
\label{fig:WrongCalibrationSettling}
\end{figure}

One additional point to consider here is the effect of using an $\alpha$ value that has been calibrated using diffusion and then applied to a grid where diffusion is not considered. The use of gravitational settling becomes problematic at higher masses. As the size of the convective envelope decreases, the surface layers can more readily be depleted, leading to models whose abundances are clearly inconsistent with observations. For this reason, the Geneva grid produced by \citet{2012A&A...541A..41M} neglects settling for stars above 1.1\ms, while the BASTI grid does not use settling even though it is included in the solar calibration \citep{2004ApJ...612..168P}. In Fig.~\ref{fig:WrongCalibrationSettling} we show tracks for both the `old' and `new' Sun, whose calibrations have been made with gravitational settling included. Alongside these, we include tracks for the same initial conditions, but without settling included. At the solar age, the models without settling are around 30\,K hotter than those with settling. By 10\,Gyr the spread in temperature has increased to around 100\,K (which is still smaller than our estimated GAIA errorbox)\footnote{We find similar results when we perform this test using {\sc mesa}.}. This presumably explains why the BASTI track is somewhat hotter than other models.

\subsection{Uncertainties in input physics}

\begin{figure}
\includegraphics[width=\columnwidth]{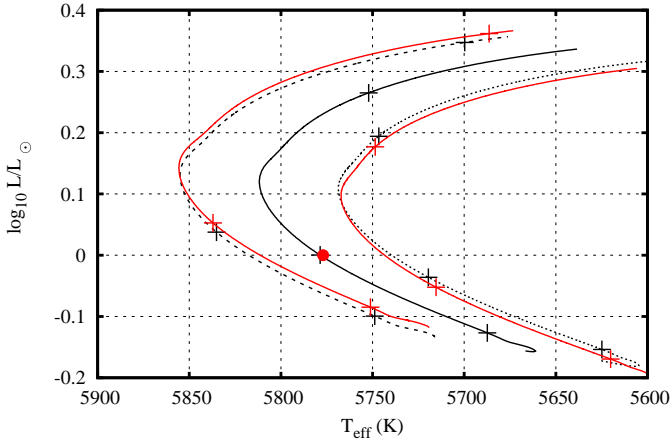}
\caption{Evolutionary tracks for a 1\ms\ model showing the effect of a small change in the hydrogen abundance. The solid line indicates the calibrated model, while the dotted and dashed lines show models where the initial hydrogen abundances has been increased and decreased by 0.01 (a change of 14\% from the calibrated value) respectively. Pluses mark ages of 500\,Myr, 4.57\,Gyr and 10\,Gyr. The red circle represents the location of the Sun. Red lines mark stellar tracks with masses of 0.98\ms\ and 1.02\ms.}
\label{fig:1smX}
\end{figure}

We now turn our attention to how small changes in the input physics can affect the stellar tracks. We use the \stars\  model with A09 abundances and settling included as a baseline. We will first deal with the effects of changes in the assumed abundances. When one applies a stellar track to a given observation, any uncertainty in the object's composition will impact the fit of the stellar model.  In Fig.~\ref{fig:1smX}, we show evolutionary tracks where we have altered the initial hydrogen abundance by $\pm0.01$. Decreasing the hydrogen abundance by 0.01 (and thus increasing the helium abundance by the same amount) makes the star hotter and more luminous. For comparison, we have also plotted tracks for stars of 0.98\ms\ and 1.02\ms. These tracks are extremely similar to our tracks with modified initial abundances. This suggests that the use of a stellar track whose hydrogen abundance is off by 0.01 (about 14\%)  can lead to a 2\% error in the determination of the stellar mass.

\begin{figure}
\includegraphics[width=\columnwidth]{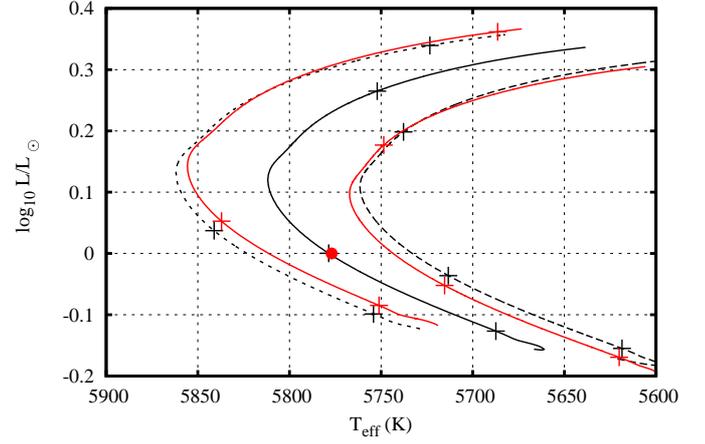}
\caption{Evolutionary tracks for a 1\ms\ model showing the effect of a small change in the total metallicity, Z. The solid line indicates the calibrated model, while the dotted and dashed lines show models where Z has been decreased and increased by 10\% respectively. Pluses mark ages of 500\,Myr, 4.57\,Gyr and 10\,Gyr. The red circle represents the location of the Sun. Red lines mark stellar tracks with masses of 0.98\ms\ and 1.02\ms.}
\label{fig:1smZ}
\end{figure}

In Fig.~\ref{fig:1smZ} we show the effects of a 10\% variation in the initial metallicity. Note for these models that we have {\it not} included any scaling of the helium abundance with Z as is typically done when constructing grids of models over large metallicity ranges. Because we have kept the initial hydrogen abundance constant, our 10\% change in metallicity corresponds to less than a 1\% change in the initial helium abundance. The effect of metallicity changes is slightly more pronounced that changes in the initial hydrogen abundance. For our three age markers, the luminosity varies by about 7-15\% while the temperature differs by no more than 2\%. Comparison of the modified metallicity tracks to our 0.98\ms\ and 1.02\ms\ (shown in grey in the figure) suggest that a 10\% error in Z amounts to an error of over 2\% in the mass determination.

\begin{figure}
\includegraphics[width=\columnwidth]{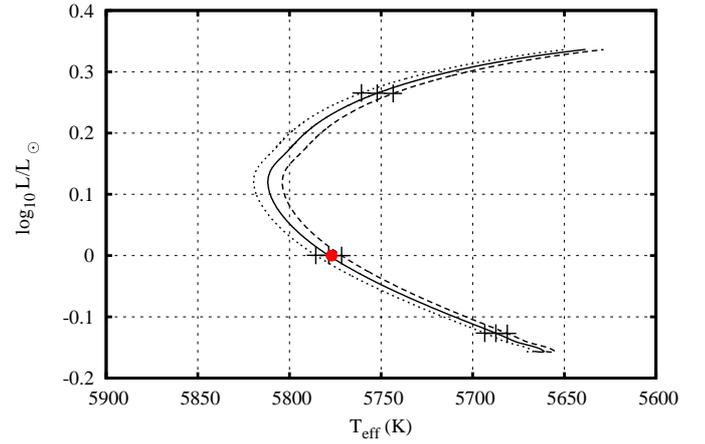}
\caption{Evolutionary tracks for a 1\ms\ model showing the effect of a small change in the mixing length parameter, $\alpha$. The solid line indicates the calibrated model, while the dotted and dashed lines show models where $\alpha$ has been set to 2.00 and 2.05 respectively. Pluses show mark ages of 500\,Myr, 4.57\,Gyr and 10\,Gyr. The red circle represents the location of the Sun.}
\label{fig:1smAlpha}
\end{figure}

In Fig.~\ref{fig:1smAlpha}, we show the effects of small changes in the mixing length coefficient $\alpha$. While the range of $\alpha$ given in Table~\ref{tab:solar} varies from 1.65 to 2.09, such a large variation in $\alpha$ would produce tracks that are clearly inconsistent with the solar calibration. We adopt a change of the order of 0.05 to represent a not unreasonable error in the solar calibration. The luminosity of each of the three age markers on our tracks is scarcely changed by the $\approx1\%$ change in $\alpha$ and they vary at the level of $\approx0.05\%$. Changes in the temperature of these 3 time points are also small, at the level of $\approx0.1\%$.

\section{3\ms\ models}

\begin{figure}
\includegraphics[width=\columnwidth]{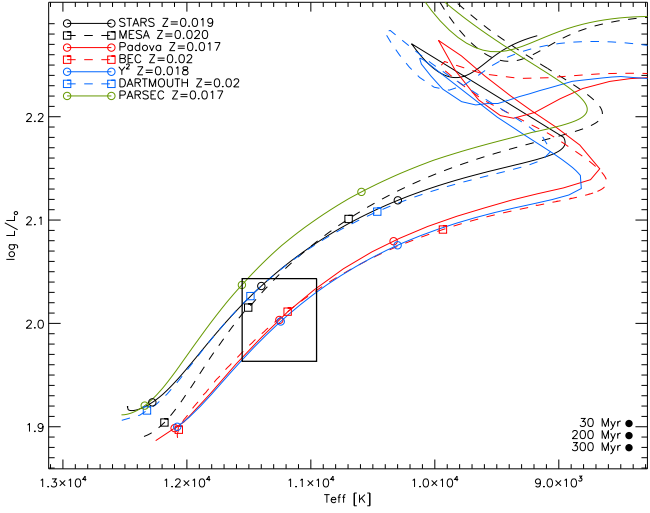}
\caption{Evolutionary tracks for a 3\ms\ model of $Z\approx0.02$. The box represents the estimated post-GAIA error on luminosity and effective temperature.}
\label{fig:3m}
\end{figure}

In Fig.~\ref{fig:3m} we compare evolutionary tracks for 3\ms\ models of solar metallicity. BASTI's canonical tracks do not go up to this mass, so we have instead plotted a 3\ms\ track from the Binary Evolution Code (BEC, see \citealt{2011A&A...530A.115B} and references therein for a description of this code). The tracks fall into two groups: the majority of codes show extremely close agreement for most of the main sequence. The Dartmouth and \stars\ tracks are hotter and more luminous than this group, though the two tracks agree extremely well with each other. We currently have no explanation for this difference. While the {\sc mesa} track initially begins close to the first group, its subsequent evolution becomes close to the \stars\ and Dartmouth tracks, though the ultimate extent of its main sequence is somewhat longer. The similarity to the \stars\ track is not surprising, given our recent recalibration of the overshooting in both codes using the same set of eclipsing binaries \citep{2015A&A...575A.117S}. The CESAM track is an outlier because it is not computed with the inclusion of convective overshooting. Considering points of fixed age, the differences between the tracks fall within the expected post-GAIA error box estimated with the same assumptions adopted in Sect. 2 with uncertainties of 0.1 mag in the bolometric correction \citep[see e.g.][]{2007A&A...470..685L} and of 300 K in effective temperature \citep[see e.g.][for a best case scenario]{2009A&A...503..945F}.

\begin{figure}
\includegraphics[width=\columnwidth]{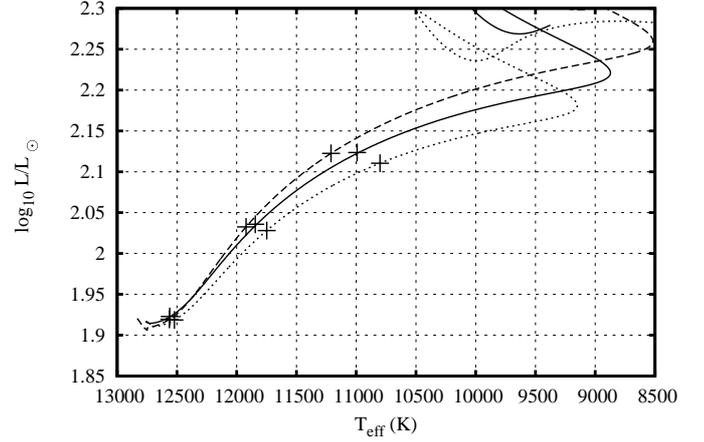}
\caption{Evolutionary tracks for a 3\ms\ model showing the effects of a small change in the overshooting parameter, \deltaov. The solid line indicates the base model with $\deltaov\ =0.15$, while the dotted and dashed lines have $\deltaov\ =0.12$ and $0.18$ respectively. Pluses mark ages of 30, 200 and 300\,Myr.}
\label{fig:3m_OS}
\end{figure}

Differences between the codes start to appear once the star approaches the end of the main sequence. This is most likely due to the treatment of convective overshooting, which affects the size of the convective core and hence the availability of hydrogen for fusion. More extensive overshooting raises the star's luminosity and the main sequence is extended (see Fig.~\ref{fig:3m_OS}). Different codes apply overshooting in different ways. Some simply extend the size of the convective region by a fixed fraction of the local pressure scale height (e.g. Dartmouth, Y$^2$). {\sc mesa} adopts an exponentially decaying mixing coefficient based on conditions close to the formal convective boundary \citep[see][for a discussion of this formalism]{1996A&A...313..497F}. The Padova group sets the boundary of the convective core at the location where convective velocities go to zero, based on the method of \citet{1981A&A...102...25B}. Finally, \stars\ applies a modification to the criterion for convective stability \citep{1997MNRAS.285..696S}, such that a region is defined to be convective unstable if $\gr > \gad - \delta$, where 
\be
\delta = {\delta_\mathrm{ov}\over 2.5 + 20\zeta +16\zeta^2}
\ee
and $\zeta$ is the ratio of radiation to gas pressure and \deltaov\ is a constant that must be determined. Despite the different formalisms (as well as the different methods to calibrate the free parameters that such formalisms necessitate), all the codes give very similar extents to the main sequence.

\begin{figure}
\includegraphics[width=\columnwidth]{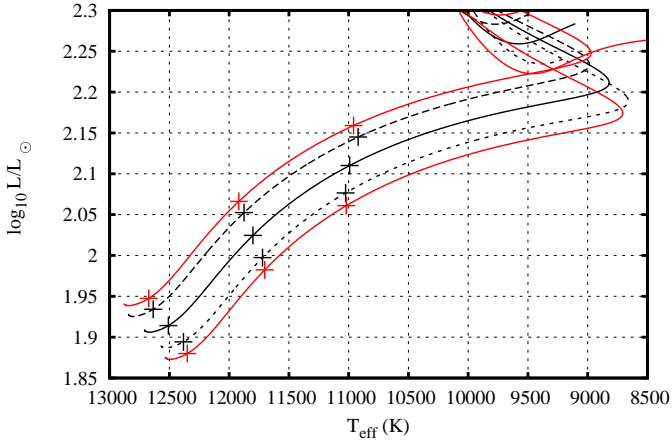}
\caption{Evolutionary tracks for a 3\ms\ model showing the effects of a small change in the initial hydrogen abundance. The dashed and dotted lines indicate models in which the initial hydrogen abundances has been reduced and increased by 0.01, respectively. Evolutionary tracks for models of 2.94 and 3.06\ms\ are shown in red. Pluses mark ages of 30, 200 and 300\,Myr.}
\label{fig:3m_X}
\end{figure}

\begin{figure}
\includegraphics[width=\columnwidth]{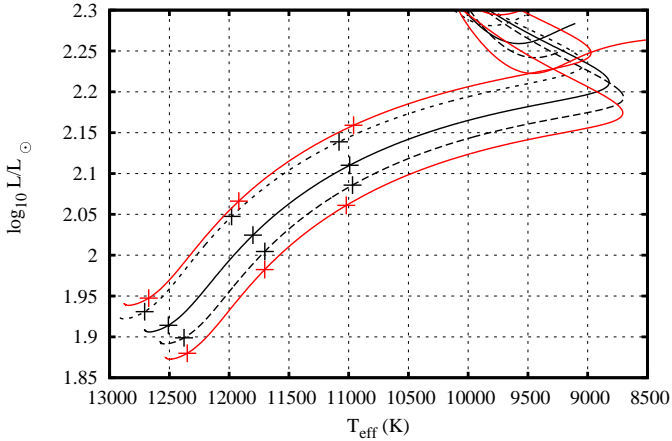}
\caption{Evolutionary tracks for a 3\ms\ model showing the effects of a small change in the initial metallicity. The dashed and dotted lines indicate models in which the initial Z has been reduced and increased by 10\%, respectively. Evolutionary tracks for models of 2.94 and 3.06\ms\ are shown in red. Pluses mark ages of 30, 200 and 300\,Myr.}
\label{fig:3m_Z}
\end{figure}

As with our 1\ms\ models, we also show the effects of small variations in both the initial hydrogen abundance (Fig.~\ref{fig:3m_X}) and the total metallicity (Fig.~\ref{fig:3m_Z}). Again, these models are computed with the {\sc stars} code. For a change of 0.01 in X, the luminosity at our three age markers (30, 200 and 300 Myr) varies between 3 and 8\% while the effective temperatures do not vary at these points by more than 1\%. The variations in hydrogen abundance are comparable to changing the mass by 1-2\%. Similar variations are found when we modify the initial metallicity by 10\%. Unlike the 1\ms\ models, the 3\ms\ models are extremely insensitive to variations in the mixing length: tracks computed with $\alpha=1.5$ are indistinguishable from those computed with $\alpha=2.025$ for the whole of the main sequence through the Hertzsprung Gap. This is because these stars do not have deep convective envelopes while on the main sequence. Note that as the stars evolve to the giant branch one would expect to see differences once the surface convective region becomes more extensive.

\section{Discussion}

Three physical inputs that we have not so far discussed and which also contribute to the uncertainties in stellar calculations are: reaction rates, opacities and the equation of state. Reaction rates are perhaps of the least concern as a source of error. The energetically important reactions for hydrogen burning (i.e. those that will affect the stellar structure) are (mostly) all well determined. As an example, the CNO cycle bottleneck reaction \el{14}{N}\pg\el{15}{O} as measured by \citet{2005EPJA...25..455I} (the recommended rate given in both the JINA REACLIB  [\citealt{2010ApJS..189..240C}] and Starlib [\citealt{2013ApJS..207...18S}] reaction rate libraries) has an uncertainty\footnote{See \citet{2013ApJS..207...18S} for a detailed discussion concerning the meaning and determination of reaction rate uncertainties.} of around 10\% in the range of temperatures pertinent to main sequence stars. Varying this rate between the upper and lower limits produces no discernible change in the evolutionary track of our 3\ms\ model computed with the \stars\ code. Note that revisions to energetically important reaction rates can have a significant impact on stellar ages. Revisions of the \el{14}{N}\pg\el{15}{O} rate by \citet{2004PhLB..591...61F} led to an increase in the predicted age of the oldest globular clusters \citep{2004A&A...420..625I}, in addition to altering the lifetimes and luminosities of advanced evolutionary phases of low-mass stars \citep{2010A&A...522A..76P}. When utilising a given set of tracks, one should always check that the reaction rates used are up-to-date.

Above a temperature of around $10^4$\,K two sets of opacities are commonly used in stellar evolution computations: the OPAL data \citep{1996ApJ...464..943I} and the OP data \citep{2005MNRAS.360..458B}. At lower temperatures, the tables of \citet{1994ApJ...437..879A} or \citet{2005ApJ...623..585F} are used. The latter are to be preferred as they are an updated and expanded version of the former. The choice of low temperature opacity tables will affect the effective temperature predicted for stars on the red giant branch, or for very low-mass stars at the beginning of the main sequence. The OPAL and OP data, which are relevant for the majority of stellar interiors considered here, generally differ at the level of 5-10\% \citep{2005MNRAS.360..458B}. This rises to around 30\% in the high temperature wing of the Z-bump, because the OP data places this bump at a slightly higher temperature \citep{2005MNRAS.360..458B}. We have used {\sc mesa} to compute calibrated 1\ms\ tracks using both OP and OPAL opacity data for solar composition of \citet{1998SSRv...85..161G}. The OP opacity is used by default in the calculations presented above. When the OPAL data is used, the parameters required are X$_0$ = 0.7095, Z$_0$ = 0.0186 and $\alpha = 1.8917$ (cf. the values in Table~\ref{tab:solar}). The evolutionary tracks are almost indistinguishable, showing a difference of no more than 3\,K at the hottest point, with the OPAL model being the cooler of the two.

The final piece of input microphysics that remains to be discussed is the choice of equation of state (EoS). Most of the codes listed employ the OPAL equation of state \citep{1996ApJ...456..902R}, whereas \stars\ employs the \citet*{1973A&A....23..325E} EoS (hereinafter EFF), with additional physics added by \citet{1995MNRAS.274..964P}, and the Dartmouth code uses an ideal gas EoS with the Debye-H\"uckel correction for masses above 0.8\ms\ \citep{1995ApJ...454..767C}. \citet{1992RMxAA..23..141D} compared the OPAL and EFF EoSs for conditions relevant to the solar hydrogen and helium ionisation zones, finding differences of the order of 1\% \citep[see also][]{1996ApJ...456..902R}. These differences were attributed to the lack of a Coulomb-pressure term \citep{1992RMxAA..23..141D}, which was subsequently added by \citet{1995MNRAS.274..964P}. The differences should presumably now be much smaller, however we cannot confirm this as we know of no recent comparisons of the various available EoSs. To test the effects of using different equations of state, we have used {\sc mesa} to compute two 3\ms\ model sequences, one using the OPAL EoS, and one using FreeEoS\footnote{The FreeEoS fortran libraries are available from \texttt{http://freeeos.sourceforge.net/}. This equation of state is a substantially modified version of the EFF EoS.}. At 30\,Myr, the difference in effective temperature between the two models is about 80\,K, with the OPAL model being the hotter of the two. 

The treatment of convection is perhaps still the Achilles' Heel of stellar evolution calculations. The mixing-length theory \citep{1958ZA.....46..108B} is still the most widely used of all the convective theories currently available (and indeed it is used in all the codes we have looked at here). From the computational viewpoint, it has the advantage of being a completely local theory so that one only needs to know about the conditions at the current mesh point. It has also worked extremely well (once properly calibrated) for a very long time. However, MLT does have more `free' parameters than just the mixing length as in the formulation of MLT one must make choices about such things as the energy exchange a blob makes with its surroundings. Not all formulations of MLT choose the same parameters and one can legitimately wonder what effect these choices may have. \citet{2008A&A...487.1075S} compared the effects of using the standard treatment of MLT used in most stellar evolution codes to the treatment commonly used in modelling white dwarf atmospheres. They showed that with appropriate calibration, the two treatments give extremely similar results. This confirms earlier work by \citet{1990ApJ...352..279P}, who also found that with appropriate calibration evolutionary tracks are uniquely defined, regardless of the MLT implementation used. Recently, \citet{2010ApJ...710.1619A} proposed a modified MLT, in which the free parameters of MLT are replaced by comparison of the relevant quantities with hydrodynamical simulations of convection. To the best of our knowledge, this has never been employed in a stellar evolution calculation though \citet{2011ASPC..445..183W} did apply it to models pulsation in asymptotic giant branch stars. 

It should be noted that one of the assumptions of MLT theory is that the mixing length obtained by calibration to the Sun is applicable to all convective situations, regardless of stellar mass, metallicity or evolutionary state. There is some observational evidence to suggest that this may not be the case \citep[e.g.][require high mixing lengths to reproduce the colours of their asymptotic giant branch stars]{2007MNRAS.378.1089M}. In addition, radiation hydrodynamics simulations of convective envelopes also suggest that the mixing length should be a function of \Teff, \logg\ and [Fe/H] \citep[e.g.][]{2011ApJ...731...78T,2014MNRAS.445.4366T,2015A&A...573A..89M}. For a more complete discussion, we refer the interested reader to the work of \citet{2015A&A...577A..60S}.

Among the alternative theories of convection is the so-called full spectrum of turbulence (FST) model of \citet{1991ApJ...370..295C}. While MLT assumes (essentially) only one large eddy, the FST model includes the contributions of turbulent eddies of all sizes. At high convective efficiencies, the FST model may produce convective fluxes that are 10 time larger than predicted by MLT \citep{1991ApJ...370..295C}. By fixing the mixing length of the FST model to be the distance to the top of the convection zone, \citet{1991ApJ...370..295C} found they were able to produce a viable solar model without recourse to free parameters. More recently, \citet{2014MNRAS.445.3592P} have also produced a parameter-free theory of convection, but it has yet to be tested in full stellar evolution models. Most recently, \citet{arnett321D} have produced a 1D convective theory based on 3D hydrodynamical simulations of convection. The implementation of such convective theories and their effect on stellar models is beyond the scope of the present work.

One area in which there is considerable variation in stellar evolution codes is the choice of boundary conditions. The surface temperature and pressure of the star are usually set close to the photosphere, either through the use of a T-$\tau$ relation, or via the use of pre-computed model atmospheres. \citet{2007AJ....134..376D} show the effects of the choice of T-$\tau$ relation on a 1\ms\ model, together with models computed using PHOENIX model atmospheres. They find that while the location of the red giant branch can be substantially affected by the choice of the relation, the main sequence is not strongly influenced. Similar results are also found by \citet{2008ApJ...675..746V}, who also included MARCS model atmospheres. In each case, when properly calibrated, the main sequence portion of the tracks are virtually identical, with substantial differences only appearing when the giant branch is reached.

\section{Conclusions}

We have attempted to quantitatively assess the theoretical uncertainties in current stellar evolution tracks. We have compared 1\ms\ tracks for eight different stellar evolution codes. We find that the spread they cover in the Hertzsprung-Russell diagram corresponds roughly to models of between 0.97-1.01\ms\ computed with the \stars\ code. 3\ms\ tracks from seven different codes have also been compared. Here, exceptional agreement is found between four of the codes, while two codes agree with each other very well despite being different from the majority group. Variations in the initial Z or X of around 10\% lead to shifts in the stellar tracks at around the 2\% level. These differences amount to a change in the initial mass of 1-2\%. Thus, if the composition of a star is not known to better than 10\%, a fit of an evolutionary track to the star's position in the HR diagram cannot give the star's mass to better than 1-2\%, even if its luminosity and effective temperature are known exactly.

\begin{acknowledgements}

We thank the anonymous referee for her/his constructive criticism of the manuscript. RJS is the recipient of a Sofja Kovalevskaja Award from the Alexander von Humboldt Foundation. LF and JCP acknowledge funding from the Alexander von Humboldt Foundation. JCP also thanks Norbert Langer for his support.

\end{acknowledgements}

\bibliographystyle{aa}
\bibliography{/Users/richardstancliffe/Work/NewBib}

\appendix

\section{{\sc mesa} solar calibration}
In table~\ref{tab:mesa_solar}, we present the details of the various solar calibrations carried out with {\sc mesa}.

\begin{sidewaystable*}
\begin{center}
\begin{tabular}{cccccccccccc}
\hline \\
Abundances  & Observables used for $\chi^2$ & $\bar{T}_{\mathrm{eff}} (K)$ & $\log{\bar{L}/L_\odot}$ & $\bar{Z}/\bar{X} $ & $\bar{He} $ & $\bar{R}_\mathrm{cz}/R_\odot $ & $\bar{c}_{\mathrm{sound}}$ & $\chi^2 $ &  [Fe/H] & Y &  $\alpha$ \\
\hline         
	GS98 & \{$L$,$T_{\mathrm{eff}}$\} & 5777 & 5.12(-10)  & 2.302(-2) & 0.2454 & - & 1.565(-3) & 1.66(-11) & 5.535(-2) & 0.2714 & 1.877\\ \\
	GS98 & \{$L$,$T_{\mathrm{eff}}$, Z/X\} & 5777 & -3.73(-7) & 2.303(-2) & 0.2454 & - & 1.563(-3) & 4.23(-4) & 5.5371(-2) & 0.2714 & 1.877 \\ \\
	GS98 & \{$L$,$T_{\mathrm{eff}}$,$R_{\mathrm{cz}}$\} & 5782 & 9.35(-5) & 2.311(-2) & 0.2456 & 7.140(-1) & 4.159(-3) & 6.48(-1) & 5.680(-2) & 0.2716 & 1.893 \\ \\
	GS98 & \{$L$,$T_{\mathrm{eff}}$,$R_{\mathrm{cz}}$,Z/X,He\} & 5781 & -7.409(-5) & 2.320(-2) & 0.2459 & 7.141(-1) & 3.652(-3) & 5.16(-1) & 5.842(-2) & 0.2719  & 1.892 \\ \\
	GS98 & \{$L$,$T_{\mathrm{eff}}$,$R_{\mathrm{cz}}$,Z/X,He,$c_s$\} & 5776 & 5.723(-5) & 2.536(-2) & 0.2526 & 7.133(-1) & 6.257(-4) & 7.01 & 9.694(-2) & 0.2788 & 1.907 \\ \\ \\ 
	
	A09 & \{$L$,$T_{\mathrm{eff}}$\} & 5777 & -3.481(-9) & 1.848(-2) & 0.2403 & - & 3.737(-3) & 7.78(-11) & -3.512(-2)  & 0.2672  & 1.790 \\ \\
	A09 & \{$L$,$T_{\mathrm{eff}}$, Z/X\} & 5777 & -1.924(-6) & 1.812(-2) & 0.2389  & - & 3.995(-3) & 4.35(-5) & -4.313(-2) & 0.2658 & 1.783 \\ \\
	A09 & \{$L$,$T_{\mathrm{eff}}$,$R_{\mathrm{cz}}$\} & 5796 & -4.033(-4) & 1.860(-2) & 0.2407 & 7.167(-1) & 1.469(-2) & 9.94 & -3.439(-2)  & 0.2670 & 1.848 \\ \\
	A09 & \{$L$,$T_{\mathrm{eff}}$,$R_{\mathrm{cz}}$,Z/X,He\} & 5797 & -9.791(-4) & 1.846(-2) & 0.2401 & 7.164(-1) & 1.600(-2) & 7.75 & -3.747(-2) & 0.2663  & 1.847\\ \\
	A09 & \{$L$,$T_{\mathrm{eff}}$,$R_{\mathrm{cz}}$,Z/X,He,$c_s$\} & 5776 & 1.171(-4) & 2.4573(-2) & 0.2610 & 7.150(-1) & 8.914(-4)  & 17.91 & 8.475(-2) & 0.2883 & 1.886 \\
\hline
\end{tabular}
\end{center}
\caption{Details of the best fit models for the various solar calibrations performed with {\sc mesa}. Numbers are given in the format $n(m)=n\times10^m$ for concision.}
\label{tab:mesa_solar}
\end{sidewaystable*}

\end{document}